\documentclass[runningheads]{llncs}
\usepackage{graphicx}
\usepackage{cite}
\usepackage{changepage}
\usepackage{subcaption}
\usepackage{enumitem}
\usepackage{bsymb}  
\usepackage{amsmath}
\usepackage {tikz}
\usetikzlibrary {positioning}
\usepackage[pdf]{graphviz}
\usepackage{array}
\usepackage{longtable}

\newcommand\changeSymbol{\texttt{+\hspace*{-.08cm}+}}
\newcommand{\chT}[2]{$\langle #1,#2\rangle$}

\newcommand{\ch}[3]{\chT{#1}{#2}\changeSymbol #3}

\newcommand{\chN}[2]{#1\texttt{+\hspace*{-.08cm}+ }#2}

\newcommand{\set}[1]{$\overline{\text{#1}}$}

\newcommand\alias{\gg}
\newcommand{\eifkw}[1]{\textbf{#1}}
\newcommand{\eif}[1]{\texttt{#1}}

\newcommand{\eifcomment}[1]{{-}{-}\textsf{#1}}

\begin{document}
\title{AutoFrame: Automatic Frame Inference for Object-Oriented Languages}
%
%
\author{Victor Rivera\inst{1,2} \and
Bertrand Meyer\inst{3,4,5}}
\authorrunning{Rivera and Meyer}
%

\institute{
Australian National University, Australia
\email{victor.rivera@anu.edu.au}
\and
Work done while the author was with Innopolis University, Russia
\and
Innopolis University, Innopolis, Russia,
\and
Schaffhausen Institute of Technology, Switzerland
\and
 Politecnico di Milano, Italy
\email{Bertrand.Meyer@inf.ethz.ch}}
\maketitle              
\begin{abstract}
Automatic program verification has made tremendous strides, but is not yet for the masses. How do we make it less painful? This article addresses one of the obstacles: the need to specify explicit ``frame clauses'', expressing what properties are left unchanged by an operation. It is fair enough to ask the would-be (human) prover to state what each operation changes, and how, but the (mechanical) prover also requires knowledge of what it does \textit{not} change. The process of specifying and verifying these properties is tedious and error-prone, and must be repeated whenever the software evolves. It is also hard to justify, since all the information about what the code changes is in the code itself.

The AutoFrame tool presented here performs this analysis entirely automatically. It applies to object-oriented programming, where the issue is compounded by aliasing: if \eif{x} is aliased to \eif{y}, any update to \eif{x.a} also affects \eif{y.a}, even though the updating instruction usually does not even mention \eif{y}. This aspect turns out to be the most delicate, and is addressed in AutoFrame by taking advantage of a companion tool, AutoAlias, which performs sound and sufficiently precise alias analysis, also in an entirely automatic way.

Some practical results of AutoFrame so far are: (1) the automatic reconstruction (in about 25 seconds on an ordinary laptop) of the exact frame clauses, a total of 169 clauses, for an  8,000-line data structures and algorithms library which was previously (with the manually written frame clauses) verified for functional correctness using a mechanical program prover; and (2) the automatic generation (in less than 4 minutes) of frame conditions for a 150,000-line graphical and GUI library. The source code of AutoFrame and these examples are available for download.

\keywords{change analysis \and frame analysis \and frame problem \and alias analysis \and Eiffel \and object-oriented programming}
\end{abstract}
\section{Introduction}

The ``frame problem''\cite{Borgida:95} is the following question: in specifying an operation, aside from stating what properties it affects and how, e.g. depositing 100 euros into a bank account increases its balance by 100, how do we avoid the tedious, error-prone and fleeting specification of what it does \textit{not} change, such as the bank account's number, the bank's address or, for that matter, the number of butterflies flapping their wings in Brazil?

Any tool for the verification of functional correctness must address this issue, since proving the correctness of an operation requires a full specification of its effect. In the \eif{deposit (sum)} example, we may expect that the programmer will write a postcondition clause stating \eif{balance = }\eifkw{old }\eif{balance + sum}, but cannot require explicit clauses \eif{owner = }\eifkw{old }\eif{owner}, \eif{account\_number = }\eifkw{old }\eif{account\_number} and so on (plus, under inheritance, new clauses for new properties introduced in descendant classes). 
The usual approach is to equip every operation with a  ``frame clause'': an exhaustive list of the properties that it is permitted to change. The operation's specification is then understood to include \eif{p = }\eifkw{old}\eif{ p} for every property \eif{p} not listed. For example the frame clause for \eif{deposit} will just list \eif{balance}. While this technique is a major improvement over the na\"ive approach of writing explicit postconditions for all non-changed properties, this article considers that it is still an undue burden on programmers, and proposes to remove it. 

Automatic program verification tools such as AutoProof \cite{Autoproof2015} and Dafny \cite{Leino:2013} successfully rely on frame clauses. But even if the specification is simpler, a human must still write it; the process is still tedious and error-prone; and it must still be repeated or at least re-checked after every program update. 

The present work proposes to avoid this process entirely by \textit{inferring} the frame clauses automatically from the program text. Indeed the implementation contains all the information needed to determine what changes. The basic rule is that an assignment \eif{x := e} changes \eif{x}. In the absence of pointers/references and aliasing, this observation would suffice for frame inference. References and paths complicate the matter: in an object-oriented (OO) language, this assignment will also change \eif{x.a}, \eif{x.a.b} etc.; in addition, it will also change \eif{y.a}, \eif{y.a.b} etc. if \eif{y} is aliased to (is a reference to) the current object (``this''). As a consequence, frame inference in an OO context, as presented in this article, fundamentally relies on \textit{alias} analysis. What makes our results possible is AutoAlias, a new alias analysis tool based on the theory of ``duality semantics'', an application of ideas from Abstract Interpretation \cite{Cousot:1977, Nielson:1999}. A companion paper \cite{Autoalias2019} describes AutoAlias.

AutoFrame is a practical tool, implemented as an addition to the EiffelStudio development environment. Its principal application so far have been to two Eiffel libraries with different scopes:
\begin{itemize}
\item 
EiffelBase 2, for a total of 169 clauses and about 8000 lines of code and 45 classes, is a formally specified library, where the specifications (contracts) define full functional correctness, which has been proved mechanically \cite{Polikarpova2015, Polikarpova2010} using the AutoProof automatic program prover. The proof, reflecting the current state of program proving technology, required manually written frame clauses. Beyond our expectations, AutoFrame infers, in a fully automatic fashion and in about 25 seconds, the \textit{exact} frame clauses of EiffelBase 2, opening a promising avenue for simplification and practicality of modern verification technology.
\item
EiffelVision 2 is a powerful and widely used (including by the EiffelStudio IDE itself for its user interface) graphical and UI library. It contains about 1141 classes and 150K LOC. AutoFrame infers all the frame clauses of EiffelVision in a little less than 4 minutes, an encouraging sign for the scalability of the approach. 
\end{itemize}

These examples, as well as the source code of AutoFrame, are available for download at \cite{AutoAlias:Impl}.

AutoFrame takes over from previous work \cite{Kogtenkov:2015, Meyer:Alias:14, BM:2010}, which had  been applied to EiffelBase+, a precursor to EiffelBase 2. Beyond an order-of-magnitude improvement in speed (the EiffelBase+ frame inference took 420 seconds), the principal difference is that the AutoFrame inference process is now entirely automatic. The previous work still required manual intervention for matching the modification of concrete \textit{class attributes} (denoting object fields), as deduced from the code, with the abstract \textit{model queries}, as used in the specification (section \ref{section:frame}). AutoFrame performs this task automatically. In particular, the exact reconstruction of EiffelBase 2 frame clauses, mentioned above, involves no manual intervention.

Some elements of this article, particularly in section \ref{frame-problem}, will at first sight look similar to the corresponding presentations in the  earlier work just cited. One of the reasons is simply to make the presentation self-contained rather than requiring the reader to go to the earlier work. More fundamentally, however, while the general approach is superficially similar, the mathematical model has been profoundly refined, and the implementation is completely new including, as noted, full automation where the previous version involved a manual step. That previous work is best viewed as a prototype for the present version.

Section \ref{section:related-work} analyzes previous work addressing automatic frame analysis. Section \ref{frame-problem} explains the framing problem and introduces the Change calculus, the basis for AutoFrame. Section \ref{section:frame} describes the AutoFrame tool and the two case studies mentioned above. Section \ref{sec:concl} discusses the potential benefits of automatic frame inference and examines an important conceptual objection, the Assertion Inference Paradox. 

It is not uncommon for articles about framing (such as \cite{Kogtenkov:2015}) to cite an extract of McCarthy and Hayes's 1969 explanation of the problem \cite{Mccarthy69somephilosophical}. Their description is so apposite as to justify that we cite it once more (with the word ``property'' replacing the more dated ``effluent'') to set the stage for the rest of the discussion: 

\begin{adjustwidth}{1cm}{1cm}
\textit{``In proving that one person could get into conversation with another, we  were obliged to add the hypothesis that if a person has a telephone he still  has it after looking up a number in the telephone book. If we had a number of actions to be performed in sequence we would have quite a number of conditions to write down that certain actions do not change the values of  certain  properties.  In  fact with n actions  and m properties we might have to write down mn such conditions.''
}
\end{adjustwidth}

\section{Related Work}
\label{section:related-work}
There is an abundant literature on the general theme of automatic code verification, and another on the aliasing issue, which plays an important role in the approach of the present work; we concentrate on references specific to frame specification, analysis and inference.

Marvin Minsky first described the frame problem in the context of artificial intelligence \cite{Minsky:1974}. Indeed it has been discussed in diverse areas including philosophy \cite{sep-frame-problem}. For the area of interest here, software verification, the classic paper is by McCarthy and Hayes, as already cited \cite{Mccarthy69somephilosophical}. 

Many verification tools provide ways to express frame properties. ``Modify'' clauses were present as early as Larch in 1993 \cite{Guttag93larch:languages}. ``Modifies'' clauses are used in a routine contract to specify which parts of the system may change as the result of the routine execution. This mechanism has been adopted by many other languages and verification tools with various names: JML \cite{Leavens:2006:PDJ:1127878.1127884} a modeling language for Java programs, uses the ``assignable'' annotation. Tools like ESC/Java2 \cite{assignable-jml:06}, a static verifier for JML-Java programs, and Krakatoa \cite{marche03krakatoa}, that translates Java programs to Coq \cite{coq-refman-09} and Why3 \cite{why:2013}, use the ``assignable'' clause to verify frame properties. Lehner et. al presented an algorithm \cite{Lehner:11} to check ``assignable'' clauses in the presence of datagroups \cite{Leino:98}; Spec\# \cite{Barnett:2004} and Dafny \cite{Leino:2013} use the ``modifies'' annotation. They use Boogie \cite{this-is-boogie-2-2},  an automatic program verifier, to check such annotations statically; Eiffel \cite{Meyer:1988} defines the \eifkw{modifies} clause and uses Autoproof \cite{Autoproof2015}, a static verifier for Eiffel based on Boogie, to prove frame conditions. In all of these approaches programmers must write the clauses manually. This differs from the work presented in the current paper as AutoFrame automatically analyzes the source code and yields the set of frame conditions.

 Rakamaric and Hu present a technique for automatically inferring frame axioms of procedures and loops using static analysis \cite{Rakamaric:08}. This work goes in the same direction as this paper, however our work is done in the context of a safe object-oriented language.

\section{The Framing Problem}
\label{frame-problem}
The frame problem is the problem of determining and verifying what properties an operation does not change. Formal specification notations include ways of defining how properties change. For example, in specification-based programming using Design by Contract  \cite{Meyer:2009:TCL, Meyer:1997:OSC}, every routine is equipped with a specification of its effect, called its contract. The contract includes a postcondition, which states the expected effect on class variables; for example an operation \eif{deposit (sum)} might have the postcondition depicted in figure \ref{ex:ensure1} (the \eifkw{ensure} clause, in the Eiffel syntax \cite{Meyer:1988}). Such a notation can also express frame conditions, as shown in figure \ref{ex:ensure2}. The impracticality of this approach is obvious, but it is still useful to list its three separate disadvantages:
\begin{enumerate}[label=(\roman*)]
\item It is tedious, since typically an operation will only change a few class variables (such as \eif{balance} for \eif{deposit}), but programmers have to write something for all the others.
\item It is fragile, since every addition of a property (a class variable) to a class will force updating all routines.
\item It does not work well with inheritance, since addition of new properties in descendant classes requires updating routines in the ancestor classes, including those that the descendants do not redefine.
\end{enumerate}

\begin{figure}[h!]
    \centering
    \begin{subfigure}[b]{0.4\textwidth}
    \[
        \begin{array}{l}
        \eifkw{class}\eif{  ACCOUNT}\\
        \hspace*{.5cm}\ldots\\
        \hspace*{.5cm}\eif{balance: INTEGER}\\
        \hspace*{.5cm}\eif{bank: BANK}\\
        \hspace*{.5cm}\eif{branch: BRANCH}\\
        \hspace*{.5cm}\ldots\\
        \hspace*{.5cm}\eif{deposit (sum: INTEGER)}\\
        \hspace*{1cm}\eifkw{do}\\
        \hspace*{1.5cm}\eif{balance := }\eif{ balance + sum}\\
    	\hspace*{1cm}\eifkw{ensure}\\
        \hspace*{1.5cm}\eif{balance = }\eifkw{old } \eif{ balance + sum}\\
        \hspace*{1cm}\eifkw{end}\\
        \eifkw{end}\\
        \end{array}
    \]
    \caption{}
    \label{ex:ensure1}
    \end{subfigure}
    \hfill 
    \begin{subfigure}[b]{0.4\textwidth}
    \[
        \begin{array}{l}
        \eifkw{class}\eif{  ACCOUNT}\\
        \hspace*{.5cm}\eif{balance: INTEGER}\\
        \hspace*{.5cm}\eif{bank: BANK}\\
        \hspace*{.5cm}\eif{branch: BRANCH}\\
        \hspace*{.5cm}\ldots\\
        \hspace*{.5cm}\eif{deposit (sum: INTEGER)}\\
        \hspace*{1cm}\eifkw{do}\\
        \hspace*{1.5cm}\eif{balance := }\eif{ balance + sum}\\
    	\hspace*{1cm}\eifkw{ensure}\\
        \hspace*{1.5cm}\eif{balance = }\eifkw{old } \eif{ balance + sum}\\
        \hspace*{1.5cm}\eif{bank = }\eifkw{old } \eif{bank}\\
        \hspace*{1.5cm}\eif{branch = }\eifkw{old } \eif{branch}\\
        \hspace*{1cm}\eifkw{end}\\
        \eifkw{end}\\
        \end{array}
      \]
    \caption{}
    \label{ex:ensure2}
    \end{subfigure}
    \caption{Using postconditions as Framing}
    \label{}
\end{figure}

To address the problem, many notations include explicit support for frame properties, such as the proposed Eiffel syntax \eifkw{only}\eif{ balance}, which states that the routine cannot change anything else than \eif{balance}. It is semantically equivalent to a whole sequence of \eif{x =}\eifkw{ old}\eif{ x} clauses, without the disadvantages. (Other possible keywords are \textbf{modify} in AutoProof \cite{Autoproof2015}, \textbf{modifies} in Dafny \cite{Leino:2013} and \textbf{assignable} in JML \cite{assignable-jml:06}.) A frame clause specifies that the operation may not modify any other properties than the ones listed, here \eif{balance}. 

The definition of such a notation must include a precise specification of its semantics, in particular its relation to inheritance. An example of frame clauses allowing mechanical program verification is the AutoProof system which, with the help of frame clauses, can automatically prove the full functional correctness of a 8000-line data structures and algorithms library, EiffelBase 2 \cite{Polikarpova2015}.

An alternative solution freeing the programmer from the obligation to specify the modifies clause (pursued in the present work), is to infer the frame conditions through automatic analysis of the routine implementation, which determines which values may change. The analysis does not adhere to modular soundness as described in \cite{Leino:2002}, however. Even though the analysis deals with the modification of extended state (it takes into consideration dynamic binding), it does need global program information to properly work. Our solution is the Change calculus and its implementation.

\subsection{Change calculus}
The Change calculus is actually a ``may-change'' calculus, which for a program instruction computes the set of locations that the instruction may change. The calculus relies on the Alias calculus \cite{Autoalias2019}. The Alias calculus is a set of rules defining the effect of executing an instruction on the aliasing that may exist between expressions. The Alias calculus builds an Alias diagram, a graph whose nodes represent possible objects and edges represent references variables. Each of the rules defining the Alias calculus gives, for an instruction \eif{$p$} of a given kind and an alias graph \eif{$G$} that holds in the initial state, the value of \eif{$G \alias p$}, the alias graph that holds after the execution of \eif{$p$}. In the Change calculus, the value of \ch{G}{c}{p}, for a program $p$, an alias graph $G$, and a change set $c$ (empty in the initial state), yields the tuple \chT{G'}{c'} containing the set of paths whose value may change as a result of executing $p$ (i.e. $c'$), and the alias graph after executing $p$ (i.e. $G'$), at some particular point in the program. Operation \set{\chT{G}{c}} yields the change set $c$. $\changeSymbol$ is an over-approximation: for soundness $c$ must include anything that changes, but conversely an expression might appear in $c$ and not change in some executions of $p$. For example, the Change calculus yields the change set $\{a,b\}$ as a result of executing the instruction  \eifkw{if}\eif{ C }\eifkw{then }\eif{a := c }\eifkw{else }\eif{b := c }\eifkw{end}. The change set expresses that the instruction may change \eif{a} and may change \eif{b}. ``May'' as in there is no implication that any particular element of this set \textit{will} change in a particular execution.

The following is the specification of the Change calculus as used for this work. The target language is a common-core subset of the mechanisms present in all modern object-oriented languages such as Java, Eiffel, C\# and C++; it is essentially the same as used for the Alias calculus \cite{Kogtenkov:2015, Meyer:Alias:14,Autoalias2019}, on which we rely.

The principal difference with actual OO languages is the ignoring of conditions: the conditional instruction is actually a non-deterministic choice, \eifkw{then} \eif{ p } \eifkw{else} \eif{ q } \eifkw{end}, without the initial ``\eifkw{if} \eif{c}'' found in ordinary languages; and similarly for the loop construct. This simplification causes a potential loss of precision, which has not, however, had visible consequences in our examples so far. For the Change calculus the problem is in fact not significant, since we have to expect that both \eif{p} and \eif{q} can be executed (we assume the command-query separation principle \cite{Meyer:1997:OSC}: asking a question should not change the answer -- functions being called in the condition are pure); otherwise the program contains dead code. Change analysis could still suffer from a loss of precision in the underlying alias analysis. The Alias calculus, however, now addresses this problem, at least in part, by including some support for conditions, as detailed in \cite{Autoalias2019}.
 
Rules of the calculus are shown in table \ref{table:change-calculus}. The rule \ch{G}{c}{p}, where $G$ is an alias graph, $c$ is a change set, initially empty and $p$ is an instruction is shown in Column \textbf{Rule}. Column \textbf{Semantics} shows the semantics of each rule. In table \ref{table:change-calculus}, $p$ and $q$ are program instructions; \eif{t}, \eif{s} and \eif{x} are path expressions. \eif{f} is a routine name and \eif{l} its actual arguments.

\begin{table}[h!]
\centering
\begin{tabular}{l|l|l}
    \hline
    \textbf{Rule Name} & \textbf{Rule} & \textbf{Semantics} \\
    \hline \hline
    
    \texttt{CC-Assg} & 
    \ch{G}{c}\eif{(t := s)} & =  \chT{G'}{c \bunion compl_{G'} (alias_{G'}(v) \bullet w)}\\
     & 
    &where $G' = G \alias $\eif{ (t := s) } and\\
    && $t = v.w \mid v \in T^*_G \wedge w \in T_G$ \\ \hline
    
    \texttt{CC-Comp} & \ch{G}{c}(p;q) &=(\chN{\ch{G}{c}{p)}}{q}\\ \hline
    
    \texttt{CC-New} & \ch{G}{c}{\eif{(}\eifkw{create }\eif{t)}} &= \chT{G\alias \eif{(}\eifkw{create }\eif{t)}}{c \bunion \{t\}}\\ \hline
    

    \texttt{CC-Cond} & \ch{G}{c}{\eif{(}$\overbrace{\eifkw{then } p \eifkw{ else } q \eifkw{ end}}^{\text{ \texttt{inst}}}$\eif{)}} &=
            \chT{G \alias\text{\texttt{inst}}}{$\set{\ch{G}{c}{p}}$ \bunion $\set{\ch{G}{c}{q}}$}
			\\\hline

      
    \texttt{CC-Loop} & \ch{G}{c}{\eif{(}$\overbrace{\eifkw{loop } p \eifkw{ end}}^{\text{\texttt{inst}}}$\eif{)}}&=
    \ch{G \alias\text{\texttt{inst}}}{c}{\eif{(}$\underbrace{p;p;\ldots}_{i \in \mathbb{N} \text{ times}}$\eif{)}}
    \\\hline
      
    \texttt{CC-UQCall} & \ch{G}{c}{\eif{(}\eifkw{call} \eif{ f(l))}}&= 
    \ch{G}{c}{$\mid \eif{f}\mid[\eif{l}:\eif{f}^\bullet]$}\\\hline
	
	
	\texttt{CC-QCall} & \ch{G}{c}{\eif{(}$\overbrace{	\eifkw{call }\eif{x.f(l)}}^{\text{\texttt{inst}}}$\eif{)}}&= 
	\eif{x}$\bullet ($\ch{G \alias \text{\texttt{inst}}}{c}{$\eifkw{call} \eif{ f(x'$\bullet$l)}$}$)$
	
	\\\hline

	\end{tabular}
\caption{The Change Calculus.}
\label{table:change-calculus}
\end{table}

    
    
    
    
    

      
	


\subsubsection{Assignments. }
The most fundamental rule for the calculus is the assignment rule as this is the instruction that defines what should change. A na\"ive (and incorrect) version of the rule is shown in figure \ref{naive:assg}. The rule states that in the assignment \eif{t := s}, variable \eif{t} is the one being changed, which is not entirely false, but it may yield an unsound result.

\begin{figure}[h!]
\centering
\begin{tabular}{l|l|l}
    \hline
    Rule Name &Rule & Semantics \\
    \hline \hline
    \texttt{CC-Assg(Naive)} & \eif{c} \changeSymbol \eif{ (t := s)} & =  \eif{$c \bunion \{t\}$}\\\hline

	\end{tabular}
\caption{Na\"ive rule for Assignment.}
\label{naive:assg}
\end{figure}

Rule in figure \ref{naive:assg} is unsound in the presence of aliasing. As an example, consider the program in figure \ref{ex:naive}. After executing instructions (1) and (2) and according to the na\"ive rule in figure \ref{naive:assg}, the change set is $\{f, f.b\}$. However, in the context of OO, variables can be references, and hence, aliasing can be present. Consider the same program in figure \ref{ex:naive} and \eif{a} and \eif{f} to be references. After the execution of instruction (1), \eif{f} is aliased to \eif{a}. This changes the result of the change set. Since \eif{f} and \eif{a} are aliased, the change set, after instruction (2) should be $\{f, f.b, a, a.b\}$.

\begin{figure}[h!]
\[
    \begin{array}{l}
    \text{(1)~~}\eif{ f := a}\\
    \text{(2)~~}\eif{ f.b := x}\\
    \end{array}
      \]
\caption{Program with aliasing}
\label{ex:naive}
\end{figure}

Rule \texttt{CC-Assg} in table \ref{table:change-calculus} copes with this situation by looking up for aliasing in $G$.  In \texttt{CC-Assg}, $alias_G (p)$ yields the set of expressions that are aliased to $p$ in $G$; $\bullet$ is the ``distributed dot'', introduced in \cite{Meyer2014}. It distributes the period of OO programming over a list, a set or a graph, for example, $x. [u,v,w] = [x.u,x.v,x.w]$; $compl_G (p)$ is the completion paths of $p$ in $G$ (defined in \cite{Autoalias2019}), i.e. all paths in $G$ starting with $p$; and $T$ is the set of all expression paths in $G$.

In words, the assignment rule works as follows. Consider the instruction \eif{t := s} being applied to a change set \eif{$c$}: First, execute the instruction \eif{t := s} on the alias graph (i.e. $G' = G \alias (t := s)$); Then, get all expressions that are aliased to $p$ in $G'$ (i.e. $alias_{G'}(p)$), where $t = p.q$. If $t$ is a class variable (not a path), then $p = $\eifkw{ Current}; Then dot-distribute it to $q$ (i.e. $alias_{G'}(p) \bullet q$); Next, compute its completion paths (i.e. $compl_{G'} (alias_{G'}(p) \bullet t)$); Finally, union the result to \eif{$c$}.

As an example, we compute the program in figure \ref{ex:naive}, with an initial alias graph as depicted in figure \ref{ex:assg:a} (it shows the \eifkw{Current} object as $\underline{n_0}$ having three class attributes $a$, $f$, and $x$), and change set $c = \emptyset$. Instruction (1) in figure \ref{ex:naive} yields the steps depicted in figure \ref{ex1:assg:steps}. (for the examples, $G_X$ represents the alias graph in figure $X$.)

Step (\ref{ex1:assg:steps}.\ref{ex1:step2}) shows the rule to be applied and its semantics. In figure \ref{ex1:assg:steps}, $G_{\text{\ref{ex:assg:b}}}$ is the  Alias graph after executing $G \alias $\eif{ (f := a)}, depicted in figure \ref{ex:assg:b}. We compute $alias_{G_{\ref{ex:assg:b}}} (\text{\eifkw{Current}}) = \{\eifkw{Current}\}$ (step \ref{ex1:assg:steps}.\ref{ex1:step3}). Next, dot-distribution $\{\eifkw{Current}\} \bullet f = \{f\}$, here $\eifkw{Current}.f = f$ according to \cite{Meyer2014} (step \ref{ex1:assg:steps}.\ref{ex1:step4}). Finally, all completion paths $compl_{G_{\text{\ref{ex:assg:b}}}} (f) = \{f, f.b\}$ (step \ref{ex1:assg:steps}.\ref{ex1:step5}). The change set is \set{\chT{G_{\ref{ex:assg:b}}}{\{f, f.b\}}}$ = \{f, f.b\}$. 
\begin{figure}[h!]
\begin{subequations}
\renewcommand{\theequation}{\roman{equation}}
\begin{align} 
 \text{\ch{G_{\ref{ex:assg:a}}}{\emptyset}{(\eif{f:=a})}}= 
& \text{\chT{G_{\ref{ex:assg:b}}}{(\emptyset \bunion compl_{G_{\ref{ex:assg:b}}}(alias_{G_{\ref{ex:assg:b}}}(\eifkw{Current}) \bullet f)}} \label{ex1:step2}\\
= & \text{\chT{G_{\ref{ex:assg:b}}}{\emptyset \bunion compl_{G_{\ref{ex:assg:b}}}(\{\eifkw{Current}\} \bullet f)}} \label{ex1:step3}\\
= & \text{\chT{G_{\ref{ex:assg:b}}}{\emptyset \bunion compl_{G_{\ref{ex:assg:b}}}(\{f\})}} \label{ex1:step4}\\
= & \text{\chT{G_{\ref{ex:assg:b}}}{\emptyset \bunion \{f, f.b\}}} 
=  \text{\chT{G_{\ref{ex:assg:b}}}{\{f, f.b\}}} \label{ex1:step5}
\end{align}
\end{subequations}
\caption{Applying \texttt{CC-Assg}}
\label{ex1:assg:steps}
\end{figure}


\begin{figure}[h!]
\centering
\begin{subfigure}[b]{0.3\textwidth}
    \begin {tikzpicture}[-latex ,auto ,node distance =2 cm  ,on grid ,
semithick , state/.style ={ circle ,top color =white ,  draw, minimum width =0.6 cm}]

    \node[state] (n0)        {\underline{$n_0$}};
    \node[state] (n1) [right =of n0] {$n_1$};
    \node[state] (n2) [right =of n1] {$n_2$};
    \node[state] (n3) [below =1.3cm of n1] {$n_3$};
    \node[state] (n4) [left =of n3] {$n_4$};
    \path (n0) edge [] node {$a$} (n1)
           (n0) edge [] node {$f$}(n3)
           (n0) edge [] node {$x$}(n4);
    \path (n1) edge [] node {$b$} (n2);
\end{tikzpicture}
\caption{$G_{\text{\ref{ex:assg:a}}}$}
\label{ex:assg:a}
\end{subfigure}
    \hspace*{1.3cm} 
\begin{subfigure}[b]{0.3\textwidth}
    \begin {tikzpicture}[-latex ,auto ,node distance =2 cm  ,on grid ,
semithick , state/.style ={ circle ,top color =white ,  draw, minimum width =0.6 cm}]

   \node[state] (n0)        {\underline{$n_0$}};
    \node[state] (n1) [right =of n0] {$n_1$};
    \node[state] (n2) [right =of n1] {$n_2$};
    \node[state] (n4) [left =of n3] {$n_4$};
    \path (n0) edge [] node {$a$} (n1)
           (n0) edge [bend right, below] node {$f$}(n1)
           (n0) edge [] node {$x$}(n4);
    \path (n1) edge [] node {$b$} (n2);
\end{tikzpicture}
\caption{$G_{\text{\ref{ex:assg:a}}} \alias ($\eif{f := a}$)$}
\label{ex:assg:b}
\end{subfigure}
\caption{Alias graphs used in figure \ref{ex1:assg:steps}}
\label{ex:assg}
\end{figure}
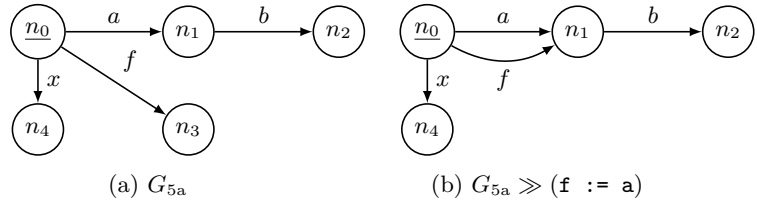

A more interesting example is to apply the calculus to instruction (2) in figure \ref{ex:naive}, with initial alias graph $G_{\text{\ref{ex:assg:b}}}$ and initial change set $c = \{f, f.b\}$. $G_{\text{\ref{ex:assg:b}}}$ shows again the \eifkw{Current} object as $\underline{n_0}$ having three class attributes $a$, $f$, and $x$, also that \eif{f} and \eif{a} are aliased. Step (\ref{ex2:assg:steps}.\ref{ex2:step2}) shows the rule to be applied and its semantics. Figure \ref{ex2:assg} depicts the effect on $G_{\ref{ex:assg:b}}$ after computing $G_{\ref{ex:assg:b}} \alias (\eif{f.b := x})$. We compute $alias_{G_{\ref{ex2:assg}}} (f) = \{a, f\}$ (step \ref{ex2:assg:steps}.\ref{ex1:step3}). Next, dot-distribution $\{a, f\} \bullet b = \{a, a.b, f, f.b\}$ (step \ref{ex2:assg:steps}.\ref{ex2:step4}). Finally, all completion paths $compl_{G_{\ref{ex2:assg}}} (\{f, f.b, a, a.b\}) = \{f, f.b, a, a.b\}$ (step \ref{ex2:assg:steps}.\ref{ex2:step5}). Step \ref{ex2:assg:steps}.\ref{ex2:step6} shows the union of the sets and the final result. The change set is \set{\chT{G_{\ref{ex2:assg}}}{\{a, a.b, f, f.b\}}}$ = \{a, a.b, f, f.b\}$.

\begin{figure}[h!]
\begin{subequations}
\renewcommand{\theequation}{\roman{equation}}
\begin{align} 
\text{\ch{G_{\ref{ex:assg:b}}}{\{f, f.b\}}{(\eif{f.b:=x})}}=& 
\text{\chT{G_{\ref{ex2:assg}}}{\{f, f.b\} \bunion compl_{G_{\ref{ex2:assg}}}(alias_{G_{\ref{ex2:assg}}}(f) \bullet b) \label{ex2:step2}}}\\
= & \text{\chT{G_{\ref{ex2:assg}}}{\{f, f.b\} \bunion compl_{G_{\ref{ex2:assg}}}(\{a, f\} \bullet b)}} \label{ex2:step3}\\
= & \text{\chT{G_{\ref{ex2:assg}}}{\{f, f.b\} \bunion compl_{G_{\ref{ex2:assg}}}(\{a, a.b, f, f.b\})}} \label{ex2:step4}\\
= & \text{\chT{G_{\ref{ex2:assg}}}{\{f, f.b\} \bunion \{a, a.b, f, f.b\}}} \label{ex2:step5}\\
= & \text{\chT{G_{\ref{ex2:assg}}}{\{a, a.b, f, f.b\}}} \label{ex2:step6}
\end{align}
\end{subequations}
\caption{Applying \texttt{CC-Assig}}
\label{ex2:assg:steps}
\end{figure}

\begin{figure}[h!]
\centering
    \begin {tikzpicture}[-latex ,auto ,node distance =2 cm  ,on grid ,
semithick , state/.style ={ circle ,top color =white ,  draw, minimum width =0.6 cm}]

   \node[state] (n0)        {\underline{$n_0$}};
    \node[state] (n1) [right =of n0] {$n_1$};\
    \node[state] (n4) [left =of n3] {$n_4$};
    \path (n0) edge [] node {$a$} (n1)
           (n0) edge [bend left, above] node {$f$}(n1)
           (n0) edge [] node {$x$}(n4);
    \path (n1) edge [] node {$b$} (n4);
\end{tikzpicture}
\caption{$G_{\text{\ref{ex:assg:b}}} \alias ($\eif{f.b := x}$)$}
\label{ex2:assg}
\end{figure}

The right-hand side of an assignment could be an expression, in such case, the alias graph is not modified. It could also be a  function (a routine returning a value), for instance $\eif{b} := \eif{get\_balance}$ where \eif{get\_balance} computes a value and returns it. The change calculus yields the change set $c = \{\eif{b}\}$ and captures the actions of the function in the alias graph $G$.

\subsubsection{Composition. } 
Rule \texttt{CC-Comp} in table \ref{table:change-calculus}
\begin{center}
    \ch{G}{c}(p;q) $=$ (\chN{\ch{G}{c}{p)}}{q}
\end{center}
    
\noindent
deals with statement composition. It states that the set of locations for $(p;q)$ is calculated by performing the analysis on the first instruction ($p$) and then the result applied on the second one ($q$). Applying the rule to the program in figure \ref{ex:naive} will follow the steps in figure \ref{ex:comp:steps}.  Step (\ref{ex:comp:steps}.\ref{comp:step1}) shows the rule to be applied and its semantics. We first compute \ch{G_{\ref{ex:assg:a}}}{\emptyset}{(\eif{f:=a})}, the result is \chT{G_{\ref{ex:assg:b}}}{\{f, f.b\}} (step \ref{ex:comp:steps}.\ref{comp:step2}) as shown in figure \ref{ex1:assg:steps}. Then, we compute \ch{G_{\ref{ex:assg:b}}}{\{f, f.b\}}{(\eif{f.b:=x})} given as a result \chT{G_{\ref{ex2:assg}}}{\{a, a.b, f, f.b\}} (step \ref{ex:comp:steps}.\ref{comp:step3}), as shown in figure \ref{ex2:assg:steps}. The change set is \set{\chT{G_{\ref{ex2:assg}}}{\{a, a.b, f, f.b\}}} $= \{a, a.b, f, f.b\}$

\begin{figure}[h!]
\begin{subequations}
\renewcommand{\theequation}{\roman{equation}}
\begin{align}
\text{\ch{G_{\ref{ex:assg:a}}}{\emptyset}{(\eif{f:=a;f.b:=x})}} = & 
\text{(\chN{\ch{G_{\ref{ex:assg:a}}}{\emptyset}{(\eif{f:=a})})}{(\eif{f.b:=x})}}\label{comp:step1}\\
=& \text{\ch{G_{\ref{ex:assg:b}}}{\{f, f.b\}}{(\eif{f.b:=x})}}\label{comp:step2}\\
=&\text{\chT{G_{\ref{ex2:assg}}}{\{a, a.b, f, f.b\}}} \label{comp:step3}
\end{align}
\end{subequations}
\caption{Applying rule \texttt{CC-Comp}}
\label{ex:comp:steps}
\end{figure}

\subsubsection{Creation.} 
Rule \texttt{CC-New} in table \ref{table:change-calculus} 
\begin{center}
    \ch{G}{c}{\eif{(}\eifkw{create }\eif{t)}} $=$ \chT{G\alias \eif{(}\eifkw{create }\eif{t)}}{c \bunion \{t\}}
\end{center}

\noindent
for creation a class variable \eif{t} is trivial: it adds a fresh node to $G$ and link it with label \eif{t} and adds \eif{t} to the change set. 

\subsubsection{Conditionals. } Rule \texttt{CC-Cond} in table \ref{table:change-calculus}

\begin{center}
\ch{G}{c}{\eif{(}$\overbrace{\eifkw{then } p \eifkw{ else } q \eifkw{ end}}^{\text{ \texttt{inst}}}$\eif{)}} $=$ \chT{G \alias\text{\texttt{inst}}}{$\set{\ch{G}{c}{p}}$ \bunion $\set{\ch{G}{c}{q}}$}
\end{center}

\noindent
deals with conditionals. The rule does not take into consideration the condition, rather treats the instruction as a non-deterministic choice. The rule assumes the command-query separation principle \cite{Meyer:1997:OSC}: asking a question should not change the answer. In other words, the rule assumes that functions being called in the condition are pure (this is not checked/forced in practice). Figure \ref{ex1:cond:steps} shows an example on how the rule is applied to the instruction \eifkw{then}\eif{ f := a}\eifkw{ else}\eif{ x := a}\eifkw{ end}. Step (\ref{ex1:cond:steps}.\ref{cond1:step1}) shows the rule to be applied and its semantics. We first compute \ch{G_{\ref{ex:condi:a}}}{\emptyset}{(\eif{f:=a})} as shown in previous examples (steps (\ref{ex1:cond:steps}.\ref{cond1:step2}) and (\ref{ex1:cond:steps}.\ref{cond1:step3})). We then compute \ch{G_{\ref{ex:condi:b}}}{\emptyset}{(\eif{x:=a})}, as shown in steps (\ref{ex1:cond:steps}.\ref{cond1:step4}) and (\ref{ex1:cond:steps}.\ref{cond1:step5}). The answer is shown in step (\ref{ex1:cond:steps}.\ref{cond1:step6}). The change set is \set{\chT{G'}{\{a, b\}}} $= \{a,b\}$.

\begin{figure}[h!]
\begin{subequations}
\renewcommand{\theequation}{\roman{equation}}
\begin{align} 
&\text{\ch{G_{\ref{ex:assg:a}}}{\emptyset}{(\eifkw{then}\text{\eif{ f:=a}\eifkw{ else}\eif{ x:=a}\eifkw{ end})}}} \nonumber\\
= &\text{\chT{G'}{$\set{\ch{G_{\ref{ex:condi:a}}}{\emptyset}{(\eif{f:=a})}}$ \bunion $\set{\ch{G_{\ref{ex:condi:b}}}{\emptyset}{(\eif{x:=a})}}$}}
\label{cond1:step1}\\
=&\text{\chT{G'}{$\set{\chT{G_{\ref{ex:condi:a}}}{\emptyset \bunion compl_{G_{\ref{ex:condi:a}}}(alias_{G_{\ref{ex:condi:a}}}(\eifkw{Current}) \bullet a)}}$ \bunion $\set{\ch{G_{\ref{ex:condi:b}}}{\emptyset}{(\eif{x:=a})}}$}}
\label{cond1:step2}\\
=&\text{\chT{G'}{$\set{\chT{G_{\ref{ex:condi:a}}}{\emptyset \bunion \{a\}}}$ \bunion $\set{\ch{G_{\ref{ex:condi:b}}}{\emptyset}{(\eif{x:=a})}}$}}
\label{cond1:step3}\\
=&\text{\chT{G'}{\{a\} \bunion $\set{\chT{G_{\ref{ex:condi:a}}}{\emptyset \bunion compl_{G_{\ref{ex:condi:b}}}(alias_{G_{\ref{ex:condi:b}}}(\eifkw{Current}) \bullet b)}}$}}
\label{cond1:step4}\\
=&\text{\chT{G'}{\{a\} \bunion $\set{\chT{G_{\ref{ex:condi:a}}}{\emptyset \bunion \{b\}}}$}}
\label{cond1:step5}\\
=&\text{\chT{G'}{\{a\} \bunion \{b\}}} = \text{\chT{G'}{\{a, b\}}}
\label{cond1:step6}
\end{align}
\end{subequations}
\caption{Applying rule \texttt{CC-Cond}}
\label{ex1:cond:steps}
\end{figure}

\begin{figure}[h!]
\centering
\begin{subfigure}[b]{0.3\textwidth}
    \begin {tikzpicture}[-latex ,auto ,node distance =2 cm  ,on grid ,
semithick , state/.style ={ circle ,top color =white ,  draw, minimum width =0.6 cm}]

    \node[state] (n0)        {\underline{$n_0$}};
    \node[state] (n1) [right =of n0] {$n_1$};
    \node[state] (n4) [below =of n0] {$n_4$};
    \path (n0) edge [] node {$a$} (n1)
           (n0) edge [bend right, below] node {$f$}(n1)
           (n0) edge [] node {$x$}(n4);
\end{tikzpicture}
\caption{$G_{\text{\ref{ex:assg:a}}} \alias (\eif{f:=a})$}
\label{ex:condi:a}
\end{subfigure}
    \hfill 
\begin{subfigure}[b]{0.3\textwidth}
    \begin {tikzpicture}[-latex ,auto ,node distance =2 cm  ,on grid ,
semithick , state/.style ={ circle ,top color =white ,  draw, minimum width =0.6 cm}]

   \node[state] (n0)        {\underline{$n_0$}};
    \node[state] (n1) [right =of n0] {$n_1$};
    \node[state] (n3) [below = of n0] {$n_3$};
    \path (n0) edge [] node {$a$} (n1)
           (n0) edge [] node {$f$}(n3)
           (n0) edge [bend right, below] node {$x$}(n1);
\end{tikzpicture}
\caption{$G_{\text{\ref{ex:assg:a}}} \alias ($\eif{x := a}$)$}
\label{ex:condi:b}
\end{subfigure}
\caption{Alias graphs used in figure \ref{ex1:cond:steps}}
\label{ex:condi}
\end{figure}
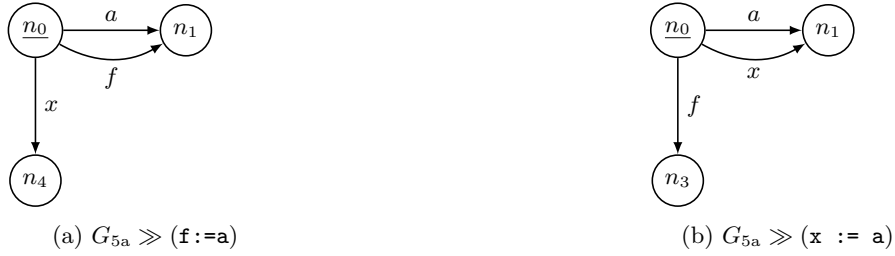

Rule \texttt{CC-Cond} is an improvement of the rule defined in \cite{Kogtenkov:2015}. Authors in \cite{Kogtenkov:2015} define the rule for conditionals to have the same effect as applying the rule \texttt{CC-Comp}. This might yield unsound results in the presence of aliasing, however. Consider the following counter example. Let $p = ($\eif{f:=a;f.b:=x}$)$ and $q = ($\eif{f.c:=d}$)$. In instruction \eifkw{then }$p$ \eifkw{else }$q$ \eifkw{end}, there is no computational path that makes the expression $a.c$ change. However, in instruction $p;q$ there is. This trivial example shows that conditional instructions should not have the same semantics as composition in the Change calculus. 

\subsubsection{Loops. } 
\eifkw{loop} $p$ \eifkw{end} is the instruction that executes $p$ any number of times including none. Rule \texttt{CC-Loop} in table \ref{table:change-calculus}

\begin{center}
    \ch{G}{c}{\eif{(}$\overbrace{\eifkw{loop } p \eifkw{ end}}^{\text{\texttt{inst}}}$\eif{)}} $=$
    \ch{G \alias\text{\texttt{inst}}}{c}{\eif{(}$\underbrace{p;p;\ldots}_{i \in \mathbb{N} \text{ times}}$\eif{)}}
\end{center}

\noindent
captures this semantics by composing $i$ times the instruction $p$, so it can produce $c$ (when $i=0$), or  \ch{G'}{c}{(p)} (when $i=1$), or \ch{G'}{c}{(p;p)} (when $i=2$) or \ch{G'}{c}{(p;p;p)} (when $i=3$) and so on. For example, notice that after the execution of the instruction \eifkw{loop } \eif{l := l.right} \eifkw{ end}, for a LinkedList \eif{l}, not just \eif{l} is allowed to change, but also \eif{l.right}, and \eif{l.right.right}, and so on. In rule \texttt{CC-Loop}, $i$ is a fixpoint. The greater the value of $i$ the more precise the analysis is, but also, the more it takes to compute a result. Empirically, we found out that $i=3$ gives a sufficient accurate result while computing it in practical times. Figure \ref{ex:loop:steps} shows the steps to apply the rule \texttt{CC-Loop} on the instruction \eifkw{loop } \eif{l := l.right} \eifkw{ end} (with $i=3$). Step (\ref{ex:loop:steps}.\ref{loop:step1}) shows the rule to be applied and its semantics. Step  (\ref{ex:loop:steps}.\ref{loop:step2}) applies rule \texttt{CC-Comp} and steps (\ref{ex:loop:steps}.\ref{loop:step3}) to (\ref{ex:loop:steps}.\ref{loop:step6}) apply the rules systematically as shown before. Step (\ref{ex:loop:steps}.\ref{loop:step7}) shows the final result. 

\begin{figure}[h!]
\begin{subequations}
\renewcommand{\theequation}{\roman{equation}}
\begin{align}
&\text{\ch{G_{\ref{example:loop:a}}}{\emptyset}{(\eifkw{loop }\eif{l := l.right} \eifkw{ end})}}\nonumber\\
=&\text{\ch{G_{\ref{example:loop:a}}}{\emptyset}{(\eif{l:=l.right; l:=l.right; l:=l.right})}}\label{loop:step1}\\
=&\text{(\chN{(\chN{\ch{G_{\ref{example:loop:a}}}{\emptyset}{(\eif{l:=l.right})})}{(\eif{l:=l.right})})}{(\eif{l:=l.right})}}\label{loop:step2}\\
&\hspace*{-1cm}=\text{(\chN{\chN{\chT{G_{\ref{example:loop:b}}}{\emptyset \bunion compl_{G_{\ref{example:loop:b}}}(alias_{G_{\ref{example:loop:b}}}(\eifkw{Current}) \bullet l)}}{(\eif{l:=l.right})})}{(\eif{l:=l.right})}}\label{loop:step3}\\
=&\text{(\chN{\chN{\chT{G_{\ref{example:loop:b}}}{\emptyset \bunion \{l, l.right\}}}{(\eif{l:=l.right})})}{(\eif{l:=l.right})}}\label{loop:step4}\\
=&\text{\chN{\chT{G_{\ref{example:loop:c}}}{\{l, l.right\} \bunion compl_{G_{\ref{example:loop:c}}}(alias_{G_{\ref{example:loop:c}}}(\eifkw{Current}) \bullet l)}}{(\eif{l:=l.right})}}\label{loop:step5}\\
=&\text{\chT{G_{\ref{example:loop:d}}}{\{l, l.right, l.right.right\} \bunion compl_{G_{\ref{example:loop:c}}}(alias_{G_{\ref{example:loop:d}}}(\eifkw{Current}) \bullet l)}}\label{loop:step6}\\
=&\text{\chT{G_{\ref{example:loop:d}}}{\{l, l.right, l.right.right, l.right.right.right\}}}\label{loop:step7}
\end{align}
\end{subequations}
\caption{Applying rule \texttt{CC-Loop}}
\label{ex:loop:steps}
\end{figure}

\begin{figure}[h!]
\centering
\begin{subfigure}[b]{0.4\textwidth}
    \begin {tikzpicture}[-latex ,auto ,node distance =1.5 cm  ,on grid ,
semithick , state/.style ={ circle ,top color =white ,  draw, minimum width =0.6 cm}]

    \node[state] (n0)        {\underline{$n_0$}};
    \node[state] (n1) [right =1.5cm of n0] {$n_1$};
    \node[state] (n2) [right =1.5cm of n1] {$n_2$};
    \node[state] (n3) [right =1.5cm of n2] {$n_3$};
    \node[state] (n4) [right =1.5cm of n3] {$n_4$};

    \path (n0) edge [] node {$l$} (n1)
           (n1) edge [] node {$right$}(n2)
           (n2) edge [] node {$right$}(n3)
           (n3) edge [] node {$right$}(n4);
\end{tikzpicture}
\caption{$G_{\ref{example:loop:a}}$}
\label{example:loop:a}
\end{subfigure}

\vspace*{.5cm}
\begin{subfigure}[b]{0.4\textwidth}
    \begin {tikzpicture}[-latex ,auto ,node distance =1.5 cm  ,on grid ,
semithick , state/.style ={ circle ,top color =white ,  draw, minimum width =0.6 cm}]

   \node[state] (n0)        {\underline{$n_0$}};
    \node[state] (n1) [right =1.5cm of n0] {$n_1$};
    \node[state] (n2) [right =1.5cm of n1] {$n_2$};
    \node[state] (n3) [right =1.5cm of n2] {$n_3$};
    \node[state] (n4) [right =1.5cm of n3] {$n_4$};

    \path (n0) edge [] node {$l$} (n1)
           (n0) edge [bend right, below] node {$l$} (n2)
           (n1) edge [] node {$right$}(n2)
           (n2) edge [] node {$right$}(n3)
           (n3) edge [] node {$right$}(n4);
\end{tikzpicture}
\caption{$G_{\ref{example:loop:a}} \alias ($\eif{l := l.right}$)$}
\label{example:loop:b}
\end{subfigure}

\vspace*{.5cm}
\begin{subfigure}[b]{0.4\textwidth}
    \begin {tikzpicture}[-latex ,auto ,node distance =1.5 cm  ,on grid ,
semithick , state/.style ={ circle ,top color =white ,  draw, minimum width =0.6 cm}]

   \node[state] (n0)        {\underline{$n_0$}};
    \node[state] (n1) [right =1.5cm of n0] {$n_1$};
    \node[state] (n2) [right =1.5cm of n1] {$n_2$};
    \node[state] (n3) [right =1.5cm of n2] {$n_3$};
    \node[state] (n4) [right =1.5cm of n3] {$n_4$};

    \path (n0) edge [] node {$l$} (n1)
          (n0) edge [bend right, below] node {$l$} (n2)
          (n0) edge [bend left] node {$l$} (n3)
           (n1) edge [] node {$right$}(n2)
           (n2) edge [] node {$right$}(n3)
           (n3) edge [] node {$right$}(n4);
\end{tikzpicture}
\caption{$G_{\ref{example:loop:b}} \alias ($\eif{l := l.right}$)$}
\label{example:loop:c}
\end{subfigure}

\vspace*{.5cm}
\begin{subfigure}[b]{0.4\textwidth}
    \begin {tikzpicture}[-latex ,auto ,node distance =1.5cm  ,on grid ,
semithick , state/.style ={ circle ,top color =white ,  draw, minimum width =0.6 cm}]

   \node[state] (n0)        {\underline{$n_0$}};
    \node[state] (n1) [right =1.5cm of n0] {$n_1$};
    \node[state] (n2) [right =1.5cm of n1] {$n_2$};
    \node[state] (n3) [right =1.5cm of n2] {$n_3$};
    \node[state] (n4) [right =1.5cm of n3] {$n_4$};

    \path (n0) edge [] node {$l$} (n1)
          (n0) edge [bend right, below] node {$l$} (n2)
          (n0) edge [bend left] node {$l$} (n3)
          (n0) edge [bend left =60] node {$l$} (n4)
           (n1) edge [] node {$right$}(n2)
           (n2) edge [] node {$right$}(n3)
           (n3) edge [] node {$right$}(n4);
\end{tikzpicture}
\caption{$G_{\ref{example:loop:c}} \alias ($\eif{l := l.right}$)$}
\label{example:loop:d}
\end{subfigure}

\caption{Alias graphs used in figure \ref{ex:loop:steps}}
\label{example:loop}
\end{figure}
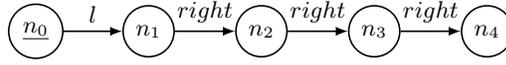
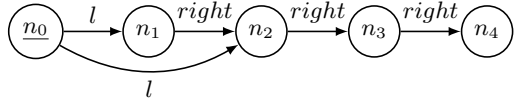
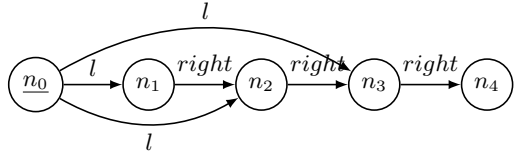
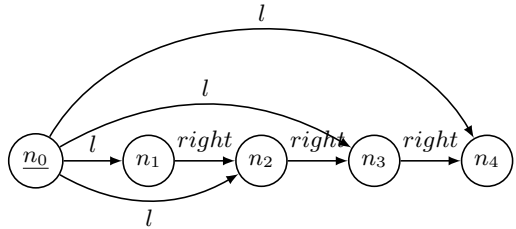

\subsubsection{Unqualified Calls. } 
Rule \texttt{CC-UQCall} in table \ref{table:change-calculus}
\begin{center}
    \ch{G}{c}{\eif{(}\eifkw{call} \eif{ f(l))}} $=$ \ch{G}{c}{$\mid \eif{f}\mid[\eif{l}:\eif{f}^\bullet]$}
\end{center}

\noindent
deals with unqualified calls (calls to routines of the Current object). \eif{l} is the list of actual arguments (arguments passed to the routine) and \eif{f$^\bullet$} is the list of formal arguments (defined by the routine) of routine \eif{f}. $\mid\eif{f}\mid$ its body. $\mid \eif{f}\mid [\eif{l}:\eif{f}^\bullet]$ substitutes all formal arguments in the body of \eif{f} with their counterpart in the actual arguments list. As an example, consider the routine defined in figure \ref{ex:setter} and the call \eif{set\_xy (a,b)}. The routine receives two arguments \eif{v} and \eif{w} of any arbitrary type \eif{T} and assigns them to variables \eif{x} and \eif{y}, respectively. The list of actual arguments is \eif{l }$ = [\eif{a}, \eif{b}]$, the list of formal arguments is \eif{set\_xy}$^\bullet = [\eif{v}, \eif{w}]$, and the substitution of all formal argument in the body of  \eif{set\_xy} with their counterpart in the actual arguments list yields $\mid\eif{set\_xy}\mid[\eif{l}:\eif{set\_xy}^\bullet] = ($\eif{x := a; y := b}$)$. If the argument passed is an expression, the substitution does not have any effect. The substitution is done to represent aliasing between formal arguments and class fields. This is important to soundly yield all locations that may be changed. If the actual argument is an expression, no aliasing exists. Assume, for example, that \eif{T} in routine \eif{set\_xy} (see figure \ref{ex:setter}) is \eif{INTEGER} and the following call \eif{set\_x (1+2, 3)}. In that case, the substitution $\mid\eif{set\_xy}\mid[\eif{l}:\eif{set\_xy}^\bullet]$ will have no effect: $\mid\eif{set\_xy}\mid[\eif{l}:\eif{set\_xy}^\bullet] = ($\eif{x := v; y := w}$)$ and the analysis will proceed normally.

\begin{figure}[h!]
\[
    \begin{array}{l}
    \eif{set\_xy (v, w: T)}\eifkw{ do}\\
    \hspace*{.5cm}\eif{x := v}\\
    \hspace*{.5cm}\eif{y := w}\\
    \hspace*{.3cm}\eifkw{end}\\
    \end{array}
      \]
\caption{Setter Routine (in Eiffel)}
\label{ex:setter}
\end{figure}

The set of steps of applying the rule to instruction \eifkw{call} \eif{set\_xy (a, b)} (routine defined in figure \ref{ex:setter}) is shown in figure \ref{ex:uqcall:steps}. Step (\ref{ex:uqcall:steps}.\ref{uqcall:step1}) shows the rule to be applied and its semantics. Step (\ref{ex:uqcall:steps}.\ref{uqcall:step2}) shows the substitution of formal arguments (here, \eif{v} and \eif{w}) for actual arguments (here, \eif{a} and \eif{b}) in the body of the routine. Steps (\ref{ex:uqcall:steps}.\ref{uqcall:step3}) and (\ref{ex:uqcall:steps}.\ref{uqcall:step4}) apply the rules systematically as shown before. Step (\ref{ex:uqcall:steps}.\ref{uqcall:step5}) shows the final result. 

\begin{figure}[h!]
\begin{subequations}
\renewcommand{\theequation}{\roman{equation}}
\begin{align}
\text{\ch{G_{\text{\ref{ex:uqcall:a}}}}{\emptyset}{\eif{(}\eifkw{call} \eif{ set\_xy(a, b))}}} =&
\text{\ch{G_{\text{\ref{ex:uqcall:a}}}}{\emptyset}{\eif{(x:=v; y:=w)}[[\eif{a}, \eif{b}]:[\eif{v}, \eif{w}]]}}\label{uqcall:step1}\\
=& \text{\ch{G_{\text{\ref{ex:uqcall:a}}}}{\emptyset}{\eif{(x:=a; y:=b)}}}\label{uqcall:step2}\\
=& \text{\chN{(\ch{G_{\text{\ref{ex:uqcall:a}}}}{\emptyset}{\eif{(x:=a)}})}{\eif{(y:=b)}}}\label{uqcall:step3}\\
=& \text{\ch{G_{\text{\ref{ex:uqcall:b}}}}{\{x\}}{\eif{(y:=b)}}}\label{uqcall:step4}\\
=& \text{\chT{G_{\text{\ref{ex:uqcall:c}}}}{\{x, y\}}}\label{uqcall:step5}
\end{align}
\end{subequations}
\caption{Applying rule \texttt{CC-UQCall}}
\label{ex:uqcall:steps}
\end{figure}

\begin{figure}[h!]
\centering
\begin{subfigure}[b]{0.3\textwidth}
    \begin {tikzpicture}[-latex ,auto ,node distance =2 cm  ,on grid ,
semithick , state/.style ={ circle ,top color =white ,  draw, minimum width =0.6 cm}]

    \node[state] (n0)        {\underline{$n_0$}};
    \node[state] (n1) [left =of n0] {$n_1$};
    \node[state] (n2) [right =of n0] {$n_2$};
    \node[state] (n3) [below =1.3cm of n1] {$n_3$};
    \node[state] (n4) [below =1.3cm of n2] {$n_4$};
    \path (n0) edge [above] node {$x$} (n1)
           (n0) edge [above] node {$y$}(n2)
           (n0) edge [below] node {$a$}(n3)
           (n0) edge [below] node {$b$}(n4);
\end{tikzpicture}
\caption{$G_{\text{\ref{ex:uqcall:a}}}$}
\label{ex:uqcall:a}
\end{subfigure}


\begin{subfigure}[b]{0.3\textwidth}
    \begin {tikzpicture}[-latex ,auto ,node distance =2 cm  ,on grid ,
semithick , state/.style ={ circle ,top color =white ,  draw, minimum width =0.6 cm}]

    \node[state] (n0)        {\underline{$n_0$}};
    \node[state] (n2) [right =of n0] {$n_2$};
    \node[state] (n3) [below =1.3cm of n1] {$n_3$};
    \node[state] (n4) [below =1.3cm of n2] {$n_4$};
    \path (n0) edge [bend right, above] node {$x$} (n3)
           (n0) edge [above] node {$y$}(n2)
           (n0) edge [below] node {$a$}(n3)
           (n0) edge [below] node {$b$}(n4)
           ;
\end{tikzpicture}
\caption{$G_{\text{\ref{ex:uqcall:a}}} \alias (\eif{x:=a})$}
\label{ex:uqcall:b}
\end{subfigure}
    \hfill 
\begin{subfigure}[b]{0.3\textwidth}
    \begin {tikzpicture}[-latex ,auto ,node distance =2 cm  ,on grid ,
semithick , state/.style ={ circle ,top color =white ,  draw, minimum width =0.6 cm}]

   \node[state] (n0)        {\underline{$n_0$}};
    \node[state] (n3) [below =1.3cm of n1] {$n_3$};
    \node[state] (n4) [below =1.3cm of n2] {$n_4$};
    \path (n0) edge [bend right, above] node {$x$} (n3)
           (n0) edge [bend left, above] node {$y$}(n4)
           (n0) edge [below] node {$a$}(n3)
           (n0) edge [below] node {$b$}(n4)
           ;
\end{tikzpicture}
\caption{$G_{\text{\ref{ex:uqcall:a}}} \alias (\eif{y:=b})$}
\label{ex:uqcall:c}
\end{subfigure}
\caption{Alias graphs used in figure \ref{ex:uqcall:steps}}
\label{ex:uqcall}
\end{figure}
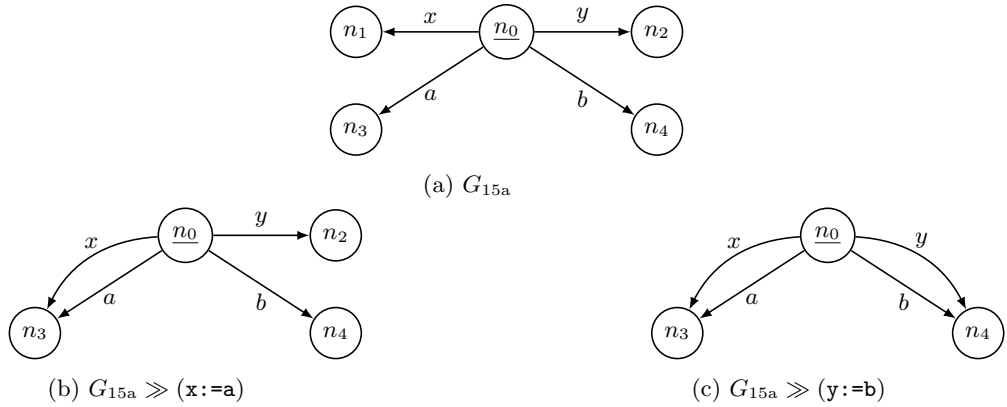

\subsubsection{Qualified Calls. } 
Rule \texttt{CC-QCall} in table \ref{table:change-calculus}

\begin{center}
    \ch{G}{c}{\eif{(}$\overbrace{	\eifkw{call }\eif{x.f(l)}}^{\text{\texttt{inst}}}$\eif{)}} $=$ 
	\eif{x}$\bullet ($\ch{G \alias \text{\texttt{inst}}}{c}{$\eifkw{call} \eif{ f(x'$\bullet$l)}$}$)$
\end{center}

\noindent
deals with qualified calls (calls to routines on a different object from Current). $\bullet$ is dot distribution \cite{Meyer2014}. It distributes the period of OO programming over lists, graphs. $x'$ (called ``negation'' of $x$) represents a back reference to the calling object, making available those variables on the target object. The set of steps of applying the rule to instruction \eifkw{call }\eif{f.set\_xy (a, b)} is shown in figure \ref{ex:qcall:steps}. Step (\ref{ex:qcall:steps}.\ref{qcall:step1}) shows the rule to be applied and its semantics. Step (\ref{ex:qcall:steps}.\ref{qcall:step2}) changes the root of the alias graph (see underline node in figure \ref{ex:qcall:b}). Instructions of the routine \eif{set\_xy} will be executed in the context defined by \eif{f}. The Alias calculus performs this operation. (\ref{ex:qcall:steps}.\ref{qcall:step3}) shows the substitution of formal arguments for actual arguments in the body of the routine. Steps (\ref{ex:qcall:steps}.\ref{qcall:step4}) to (\ref{ex:qcall:steps}.\ref{qcall:step6}) apply the rules systematically as shown before. Step (\ref{ex:qcall:steps}.\ref{qcall:step7}) performs dot distribution: it restores the root of the alias graph (as to perform operations back in the target of the call, see figure \ref{ex:qcall:e}) and distributes to the change set (as to represent that those locations that may change are in another object).

\begin{figure}[h!]
\begin{subequations}
\renewcommand{\theequation}{\roman{equation}}
\begin{align}
&\text{\ch{G_{\text{\ref{ex:qcall:a}}}}{\emptyset}{\eif{(}\eifkw{call }\eif{f.set\_xy(a, b))}}}\nonumber\\
& = \text{\eif{f}$\bullet ($\ch{G_{\text{\ref{ex:qcall:a}}}}{\emptyset}{\eif{(}\eifkw{call} \eif{ set\_xy (f'.a, f'.b))}}$)$}\label{qcall:step1}\\
& = \text{\eif{f}$\bullet ($\ch{G_{\text{\ref{ex:qcall:b}}}}{\emptyset}{\eif{(x:=v; y:=w)[[\eif{f'.a}, \eif{f'.b}]:[\eif{v}, \eif{w}]]}}$)$}\label{qcall:step2}\\
& = \text{\eif{f}$\bullet ($\ch{G_{\text{\ref{ex:qcall:b}}}}{\emptyset}{\eif{(x:=f'.a; y:=f'.b)}}$)$}\label{qcall:step3}\\
& = \text{\eif{f}$\bullet ($\chN{(\ch{G_{\text{\ref{ex:qcall:b}}}}{\emptyset}{\eif{(x:=f'.a)})}}
{\eif{(y:=f'.b)}$)$}}\label{qcall:step4}\\
& = \text{\eif{f}$\bullet ($\chN{\chT{G_{\text{\ref{ex:qcall:c}}}}{\{x\}}}{\eif{(y:=f'.b)}$)$}}\label{qcall:step5}\\
& = \text{\eif{f}$\bullet ($\chT{G_{\text{\ref{ex:qcall:d}}}}{\{x, y\}}$)$}\label{qcall:step6}\\
& = \text{\chT{G_{\text{\ref{ex:qcall:e}}}}{\{f.x, f.y\}}}\label{qcall:step7}
\end{align}
\end{subequations}
\caption{Applying rule \texttt{CC-QCall}}
\label{ex:qcall:steps}
\end{figure}

\begin{figure}[h!]
\centering
\begin{subfigure}[b]{0.3\textwidth}
    \begin {tikzpicture}[-latex ,auto ,node distance =2 cm  ,on grid ,
semithick , state/.style ={ circle ,top color =white ,  draw, minimum width =0.6 cm}]

    \node[state] (n0)        {\underline{$n_0$}};
    \node[state] (n1) [left =of n0] {$n_1$};
    \node[state] (n2) [right =of n0] {$n_2$};
    \node[state] (n3) [below =1.3cm of n0] {$n_3$};
    \node[state] (n4) [left =of n3] {$n_4$};
    \node[state] (n5) [right =of n3] {$n_5$};
    \path (n0) edge [above] node {$a$} (n1)
           (n0) edge [above] node {$b$}(n2)
           (n0) edge [] node {$f$}(n3);
    \path (n3) edge [below] node {$x$}(n4)
          (n3) edge [below] node {$y$}(n5);
\end{tikzpicture}
\caption{$G_{\text{\ref{ex:qcall:a}}}$}
\label{ex:qcall:a}
\end{subfigure}
    \hfill 
\begin{subfigure}[b]{0.3\textwidth}
    \begin {tikzpicture}[-latex ,auto ,node distance =2 cm  ,on grid ,
semithick , state/.style ={ circle ,top color =white ,  draw, minimum width =0.6 cm}]

  \node[state] (n0)        {$n_0$};
    \node[state] (n1) [left =of n0] {$n_1$};
    \node[state] (n2) [right =of n0] {$n_2$};
    \node[state] (n3) [below =1.3cm of n0] {\underline{$n_3$}};
    \node[state] (n4) [left =of n3] {$n_4$};
    \node[state] (n5) [right =of n3] {$n_5$};
    \path (n0) edge [above] node {$a$} (n1)
           (n0) edge [above] node {$b$}(n2)
           (n0) edge [] node {$f$}(n3);
    \path (n3) edge [below] node {$x$}(n4)
          (n3) edge [below] node {$y$}(n5)
          (n3) edge [bend left, red, dotted] node {$f'$}(n0);
\end{tikzpicture}
\caption{$f' \bullet G_{\text{\ref{ex:qcall:a}}}$}
\label{ex:qcall:b}
\end{subfigure}

\begin{subfigure}[b]{0.3\textwidth}
    \begin {tikzpicture}[-latex ,auto ,node distance =2 cm  ,on grid ,
semithick , state/.style ={ circle ,top color =white ,  draw, minimum width =0.6 cm}]

    \node[state] (n0)        {$n_0$};
    \node[state] (n1) [left =of n0] {$n_1$};
    \node[state] (n2) [right =of n0] {$n_2$};
    \node[state] (n3) [below =1.3cm of n0] {\underline{$n_3$}};
    \node[state] (n5) [right =of n3] {$n_5$};
    \path (n0) edge [above] node {$a$} (n1)
           (n0) edge [above] node {$b$}(n2)
           (n0) edge [] node {$f$}(n3);
    \path (n3) edge [below, bend left] node {$x$}(n1)
          (n3) edge [below] node {$y$}(n5)
          (n3) edge [bend left, red, dotted] node {$f'$}(n0);
\end{tikzpicture}
\caption{$G_{\text{\ref{ex:qcall:b}}} \alias (\eif{x:=a})$}
\label{ex:qcall:c}
\end{subfigure}
    \hfill 
\begin{subfigure}[b]{0.3\textwidth}
    \begin {tikzpicture}[-latex ,auto ,node distance =2 cm  ,on grid ,
semithick , state/.style ={ circle ,top color =white ,  draw, minimum width =0.6 cm}]

    \node[state] (n0)        {$n_0$};
    \node[state] (n1) [left =of n0] {$n_1$};
    \node[state] (n2) [right =of n0] {$n_2$};
    \node[state] (n3) [below =1.3cm of n0] {\underline{$n_3$}};
    \path (n0) edge [above] node {$a$} (n1)
           (n0) edge [above] node {$b$}(n2)
           (n0) edge [] node {$f$}(n3);
    \path (n3) edge [below, bend left] node {$x$}(n1)
          (n3) edge [bend right, below] node {$y$}(n2)
          (n3) edge [bend left, red, dotted] node {$f'$}(n0);
\end{tikzpicture}
\caption{$G_{\text{\ref{ex:qcall:c}}} \alias (\eif{y:=b})$}
\label{ex:qcall:d}
\end{subfigure}

\begin{subfigure}[b]{0.3\textwidth}
    \begin {tikzpicture}[-latex ,auto ,node distance =2 cm  ,on grid ,
semithick , state/.style ={ circle ,top color =white ,  draw, minimum width =0.6 cm}]

    \node[state] (n0)        {\underline{$n_0$}};
    \node[state] (n1) [left =of n0] {$n_1$};
    \node[state] (n2) [right =of n0] {$n_2$};
    \node[state] (n3) [below =1.3cm of n0] {$n_3$};
    \path (n0) edge [above] node {$a$} (n1)
           (n0) edge [above] node {$b$}(n2)
           (n0) edge [] node {$f$}(n3);
    \path (n3) edge [below, bend left] node {$x$}(n1)
          (n3) edge [bend right, below] node {$y$}(n2);
\end{tikzpicture}
\caption{$\eif{f} \bullet G_{\text{\ref{ex:qcall:d}}}$}
\label{ex:qcall:e}
\end{subfigure}
\caption{Alias graphs used in figure \ref{ex:qcall:steps}}
\label{ex:qcall}
\end{figure}
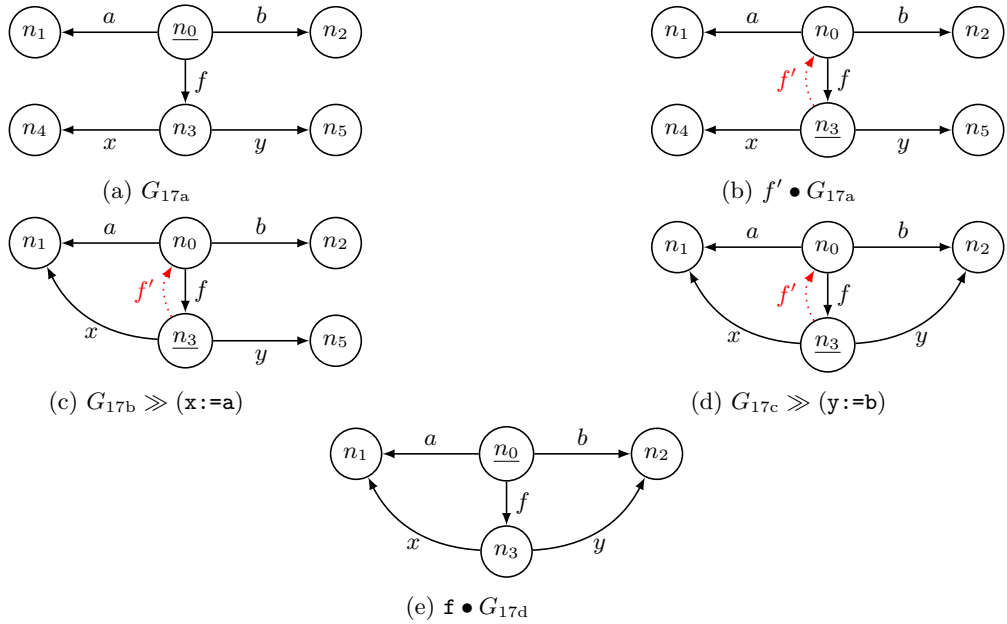

\paragraph{Dynamic Binding}
Rule \texttt{CC-QCall} needs to be extended to be able to deal with Dynamic binding. Dynamic binding might exist in the presence of Inheritance, a mechanism that enables users to create `is-a' relations between different classes: considering \eif{A} and \eif{B} as types, if \eif{B} inherits from \eif{A}, whenever an instance of \eif{A} is required, an instance of \eif{B} will be acceptable. This mechanism enables entities to be polymorphic at run-time, that is, a dynamic entity's type might differ from its static type. Dynamic binding is the property that any execution of a feature call will use the version of the feature best adapted to the type of the target object. This can only be determined at run-time. Since the Chance calculus analyzes the source code statically, it is not possible to know upfront what the appropriate type of a specific entity is. Consider, as an example, classes depicted in figure \ref{ex:dynBin}. Class \eif{T1}, in figure \ref{ex:dynBin:a}, defines a class field  \eif{c}. Class \eif{T2}, in figure \ref{ex:dynBin:b}, inherits from class \eif{T1} (by using the keyword \eifkw{inherit}), it also gives a redefinition of routine \eif{set} (indicated by the keyword \eifkw{redefine}), and defines a new class variable \eif{b}. 

\begin{figure}[h!]
    \centering
    \begin{subfigure}[b]{0.3\textwidth}
    \[
        \begin{array}{l}
    	\eifkw{class}\eif{ T1}\\
		\eifkw{feature}\\
        \hspace*{.5cm}\eif{c: T}\\
        \hspace*{.5cm}\eif{set (arg: T) }\eifkw{ do}\\
		\hspace*{1cm}\eif{ c := arg}\\
		\hspace*{.5cm}\eifkw{end}\\
        \eifkw{end}\\
        
        \end{array}
    \]
    \caption{}
    \label{ex:dynBin:a}
    \end{subfigure}
    \hfill 
    \begin{subfigure}[b]{0.5\textwidth}
    \[
        \begin{array}{l}
    	\eifkw{class}\eif{ T2}\\
    	\eifkw{inherit}\eif{ T1 }\eifkw{ redefine }\eif{ set }\eifkw{ end}\\
		\eifkw{feature}\\
		\hspace*{.5cm}\eif{b: T}\\
        \hspace*{.5cm}\eif{set (arg: T) }\eifkw{ do}\\
		\hspace*{1cm}\eif{ b := arg}\\
		\hspace*{.5cm}\eifkw{end}\\
        \eifkw{end}\\
        
        \end{array}
      \]
    \caption{}
    \label{ex:dynBin:b}
    \end{subfigure}
    
    \begin{subfigure}[b]{0.5\textwidth}
    \[
        \begin{array}{l}
    	\eifkw{class}\eif{ B}\\
    	\eifkw{feature}\\
    	\hspace*{.5cm}\eif{t, a: T1}\\
        \hspace*{.5cm}\eif{call\_set}\eifkw{ do}\\
		\hspace*{1cm}\eif{ t.set (a)}\\
		\hspace*{.5cm}\eifkw{end}\\
        \eifkw{end}\\
        
        \end{array}
      \]
    \caption{}
    \label{ex:dynBin:call}
    \end{subfigure}
    \caption{Dynamic Binding example}
    \label{ex:dynBin}
\end{figure}

The question is what should the change set contain after the execution of routine \eif{call\_set} in class \eif{B}? This cannot be inferred statically as the object attached to \eif{t} in call \eif{t.set (a)} might be of type \eif{T1} or \eif{T2}. Using the static type of \eif{t} might result in a unsound change set: analysing feature \eif{set} in class \eif{T1} will yield the change set $\{t.c\}$. This is not sound as one might call the routine as follows:

\noindent
\ldots\\
\eif{w: T2}\\
\ldots\\
\eif{t := w}\\
\eif{call\_set }\\
\ldots

In this case, the change set should contain $\{t.b\}$. The correct answer, for this example, is that the change set should contain all locations that may be changed by routine \eif{set} defined in class \eif{T1} as well as any routine \eif{set} redefined in any heir of \eif{T1}, here \eif{T2}. So the change set should be $\{t.c, t.b\}$. The mechanism to achieving that goal is to treat the instruction as a conditional where the branches of the condition are the different heirs of the static type of the target being analysed. For this example it would be:

\noindent
\eifkw{then}\\
\hspace*{.5cm}t.set (a) \textit{ -- considering \eif{t} attached to \eif{T1}}\\
\eifkw{else}\\
\hspace*{.5cm}t.set (a) \textit{ -- considering \eif{t} attached to \eif{T2}}\\
\eifkw{end}

It will consider as many branches as heirs of \eif{T1} exists that redefine the feature call. This mechanism introduces imprecision but retains soundness.

\section{AutoFrame: a tool for automatically synthesizing frame conditions}
\label{section:frame}
AutoFrame is an implementation of the Change calculus described in the previous section. It relies on the implementation of the Alias calculus (as shown, for instance, by rule \texttt{CC-Assg}).  AutoFrame statically analyzes the source code of a routine and yields the set of class attributes that are allowed to be changed. Such a set is a suggestion of Frame Conditions to programmers. The tool is implemented in Eiffel and integrated in EiffelStudio. The tool generates Eiffel \eifkw{modify} clauses (as shown in figure \ref{fig:class}). This modify clauses can be used by Autoproof to prove the correctness of a class. Sources of the tool are available in \cite{AutoAlias:Impl} and results can be checked in \cite{AutoAliasAutoFrame:results}.

\subsection{Using AutoFrame}
\label{subsection:using}
AutoFrame has been used on two libraries for Eiffel, EiffelVision and EiffelBase. The tool successfully yielded the Change Set for each routine in the libraries as a collection of expressions in a \eifkw{modify} clause.

\subsubsection{EiffelVision.} 
EiffelVision 2 is the basic library for building graphical and GUI (Graphical User Interface) applications in the Eiffel programming language. It has around 150K Lines of Code (LOC) and 1141 classes. AutoFrame automatically suggested Frame Conditions for all routines in the library. AutoFrame performed the analysis in around 232 seconds. 

The EiffelVision library is not equipped with modify clauses (as the EiffelBase 2 library described in Section \ref{subsection:eiffelbas2}). Hence, we cannot be sure about the soundness of the result. We manually checked (randomly) some of the analyzed features finding no inconsistencies. 

\subsubsection{EiffelBase 2.}
\label{subsection:eiffelbas2}
EiffelBase 2 is a formally specified and verified library \cite{Polikarpova:2013} for Eiffel. It contains a set of classes that implement common data structures and algorithms. One of the main advantages to work with this library is that all classes are equipped with contracts (pre, postconditions and class invariants) that specify classes' behavior. The specifications of the classes are fully verified against its implementation. 

The specification style of the library relies on mathematical ``model queries'' \cite{Polikarpova2010}. Each class in EiffelBase 2 declares its abstract state through a set of model attributes. Figure \ref{fig:class} is an excerpt of class \eif{V\_LINKED\_LIST}, a class that implements linked lists in Eiffel. Its model attribute (after the keyword \eifkw{model}) is a sequence of list elements: \eif{sequence}. Its type \eif{MML\_SEQUENCE} is from a Mathematical Model Library (MML). Each command (method) with observable side effects, such as \eif{extend\_back} in figure \ref{fig:class}, defines a ``modifies'' clause that lists the model attributes that are allowed to be changed by the command. The library is equipped with a total of 169 modify clauses. 

\begin{figure}[h!]
{\scriptsize
\[
\begin{array}{p{14cm}}
\eifkw{class}\eif{ V\_LINKED\_LIST [G]}\eifkw{ inherit }\eif{V\_LIST [G]}\\
\eifkw{model}\eif{: sequence}\\
\eifkw{feature}\eif{ \{}\eifkw{public}\eif{\}}\\
\hspace*{.5cm}\eifkw{ghost}\eif{ sequence: MML\_SEQUENCE [G]}\\
\hspace*{.5cm}\eif{extend\_back (v: G) }\eifcomment{Insert `v' at the back.}\\
\hspace*{1cm}\eifkw{modify\_model: }\eif{sequence}\\
\hspace*{1cm}\eifkw{local}\eif{ cell: V\_LINKABLE [G]}\\
\hspace*{1cm}\eifkw{do}\\
\hspace*{1.5cm}\eifkw{create}\eif{ cell.put (v)}\\
\hspace*{1.5cm}\eifkw{if}\eif{ first\_cell = }\eifkw{Void then}\\
\hspace*{2cm}\eif{first\_cell := cell}\\
\hspace*{1.5cm}\eifkw{else}\\
\hspace*{2cm}\eif{last\_cell.put\_right (cell)}\\
\hspace*{1.5cm}\eifkw{end}\\
\hspace*{1.5cm}\eif{last\_cell := cell}\\
\hspace*{1.5cm}\eif{cells := cells \& cell}\\
\hspace*{1.5cm}\eif{sequence := sequence \& v}\\
\hspace*{1cm}\eifkw{ensure}\eif{ sequence = }\eifkw{old}\eif{ sequence + (v)}\\
\hspace*{1cm}\eifkw{end}\\
\eifkw{feature}\eif{ \{}\eifkw{private}\eif{\} }\\
\hspace*{.5cm}\eif{first\_cell: V\_LINKABLE [G] }\eifcomment{First cell of the list.}\\
\hspace*{.5cm}\eif{last\_cell: V\_LINKABLE [G] }\eifcomment{Last cell of the list.}\\
\hspace*{.5cm}\eifkw{ghost}\eif{ cells: MML\_SEQUENCE [LINKABLE [G]]}\\
\eifkw{invariant}\\
\hspace*{.5cm}\eif{cells\_domain: sequence.count = cells.count}\\
\hspace*{.5cm}\eif{first\_cell\_empty: cells.is\_empty = (first\_cell = }\eifkw{Void}\eif{)}\\
\hspace*{.5cm}\eif{last\_cell\_empty: cells.is\_empty = (last\_cell = }\eifkw{Void}\eif{)}\\
\hspace*{.5cm}\eif{cells\_exist: cells.non\_void}\\
\hspace*{.5cm}\eif{sequence\_implementation: }\eifkw{across}\eif{ 1 |..| cells.count }\eifkw{as}\eif{ i }\eifkw{all}\eif{ sequence [i.item] = cells [i.item].item }\eifkw{end}\\
\hspace*{.5cm}\eif{cells\_linked: is\_linked (cells)}\\
\hspace*{.5cm}\eif{cells\_first: cells.count > 0}\eifkw{ implies}\eif{ first\_cell = cells.first}\\
\hspace*{.5cm}\eif{cells\_last: cells.count > 0 }\eifkw{implies}\eif{ last\_cell = cells.last }\eifkw{and then}\eif{ last\_cell.right=}\eifkw{Void}\\
\eifkw{end}\\
\end{array}
\]
} 
\caption{Excerpt from EiffelBase 2 class \eif{V\_LINKED\_LIST}}
\label{fig:class}
\end{figure} 

As a validation step of AutoFrame, we performed the Frame analysis on the Eiffel Base 2 library to automatically generate the change set, the modify clauses, of each routine. Since the library already contains those clauses, we were interested on whether Autoframe yields a subset, superset or the same set of change locations as the ones already present in the library. AutoFrame was able to suggest frame conditions to all features of the library. It automatically generated the 169 modify clauses. The tool analyzed a total of 45 classes containing 513 features (around 8K LOC). It does so in around 25 seconds. This timing outperforms the previous relation-based implementation of Change calculus \cite{Kogtenkov:2015} that takes around 420 seconds to analyze a precursor to EiffelBase 2 library. 

\begin{figure}
\centering
\includegraphics[width=4in]{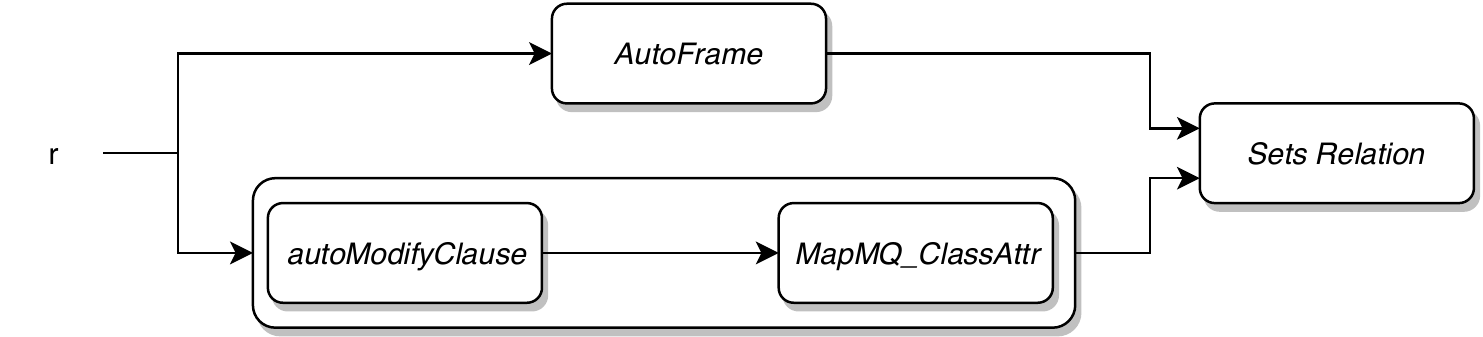}
\caption{AutoFrame validation}
\label{fig:val1}
\end{figure}

Figure \ref{fig:val1} depicts the general execution of the validation. For all routines \eif{r} in EiffelBase 2, the routine goes to two processes, each process yields a set of class attributes, the validation consists in checking whether both sets are the same. In the first process (\textit{AutoFrame} in figure \ref{fig:val1}), AutoFrame automatically infers the modify clause of routine \eif{r}; in the second process, routine \eif{r} is passed to \textit{autoModifyClause}, a helper tool that automatically retrieves all information from the ``modify'' clause that comes with EiffelBase (e.g. \eif{sequence} in routine \eif{extend\_back} in figure \ref{fig:class}). The specification style of the library relies on mathematical ``model queries'', hence, only model queries are listed in the clause, and not program attributes directly. In order to compare the results of each process, it is necessary to map model queries to class attributes. This job is performed by \textit{MapMQ\_ClassAttr}, another helper tool that automatically maps model queries to class attributes by performing data dependency analysis. Finally, both resulted sets are passed to \textit{Sets Relation} component that yields the relation between both sets: \textit{subset}, \textit{superset} or \textit{equals}.

AutoFrame was able to suggest exactly the same set of class attributes as the ones listed by the modify clause of the library. EiffelBase 2 has been formally verified so we were not expecting to find discrepancies, what it is interesting is that AutoFrame is able to infer all Frame Conditions automatically by analyzing the source code.

\subsection{Verification of Frame conditions}

We performed another validation step on the Frame conditions automatically infered by AutoFrame. We were interested on verifying that the set of locations listed by AutoFrame is indeed the set of locations allowed to be changed. This can be achieved  if the code being analyzed is equipped with contracts, in particular with post-conditions. An informal review performed on publicly available Java Modeling Language (JML) \cite{Leavens:2006:PDJ:1127878.1127884} code revealed that in practice class attributes mentioned in a modifies (``assignable'' in JML) clause for a command also appear in the postcondition of that command. In other words, it seems that whenever JML programmers state that something can be modified they also say how it will be modified. They do not necessarily say it in exact terms, as in \eif{q = some\_value}, but may just state \eif{some\_property (q)}. Either way, however, the postcondition names \eif{q}. It then seems a waste of effort to require writing a special clause listing such class attributes. Since EiffelBase 2 is equipped with postconditions, we took advantange of it. 

The verification of Frame conditions, having postconditions, can then be performed in two steps. This first step is to analyze the code of a routine \eif{r} and list all class attributes that \eif{r} is modifying. This step is called Frame Implementation Inference (\texttt{FII}) and is the one carried out by AutoFrame. The second step is to analyze the postcondition of \eif{r} to list all class attributes it mentions. This step is called Frame Specification Inference (\texttt{FSI}).  The analysis of the postcondition should only consider class attributes outside of an \eifkw{old} clause: a clause \eif{q1 = some\_function (}\eifkw{old}\eif{ q2)} indicates that \eif{q1} can be modified, but says nothing about \eif{q2}. 

The verification of Frame Conditions becomes trivial having the set of class attributes being modified by the routine (i.e. \texttt{FII}) and the set of class attributes being named in the postcondition (i.e. \texttt{FSI}), the Frame Condition holds if

$$
    \texttt{FII} \subseteq \texttt{FSI}.
$$

\begin{figure}
\centering
\includegraphics[width=4in]{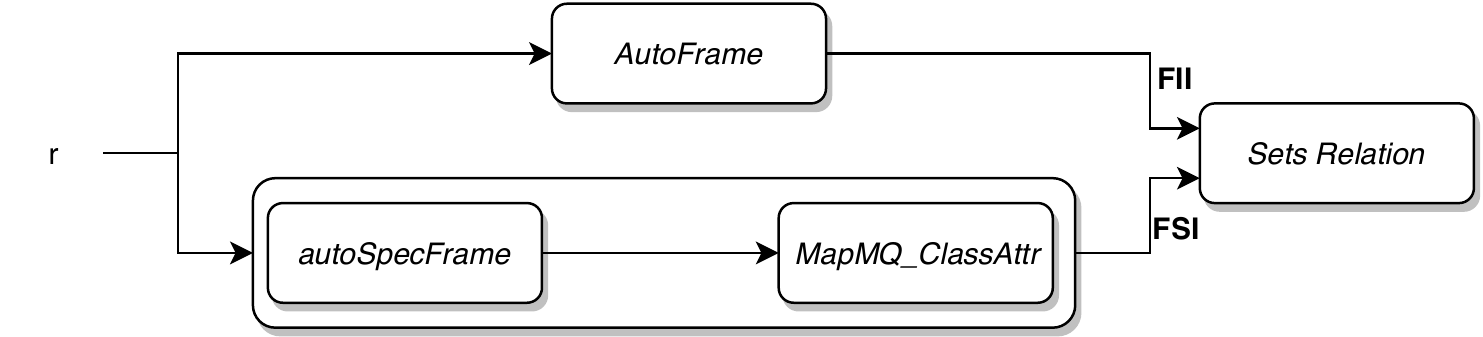}
\caption{Frame Condition verification}
\label{fig:ver}
\end{figure}

We conducted a Frame Condition Verification on the EiffelBase 2 library. Figure \ref{fig:val1} depicts the general execution of the process. For all routines \eif{r} in EiffelBase 2, the routine goes to two processes, each process yields a set of class attributes, the verification consists in checking whether the property $\texttt{FII} \subseteq \texttt{FSI}$ holds. In the first process (\textit{AutoFrame} in figure \ref{fig:ver}), AutoFrame automatically infers the modify clause of routine \eif{r}; in the second process, routine \eif{r} is passed to \textit{autoSpecFrame}, a helper tool that statically analyzes the postcondition of \eif{r} and lists all model queries being named on it (e.g. \eif{sequence} in the postcondion, after the \eifkw{ensure} part, of routine \eif{extend\_back} in figure \ref{fig:class}). As mentioned before, the specification style of the library relies on mathematical ``model queries'', hence, only model queries are listed in the clause, and not program attributes directly. In order to compare the results of each process, it is necessary to map model queries to class attributes. This job is performed by \textit{MapMQ\_ClassAttr}, another helper tool that automatically maps model queries to class attributes by performing data dependency analysis. These two process yield sets \texttt{FII} and \texttt{FSI}. Finally, both sets \texttt{FII} and \texttt{FSI} are passed to the \textit{Sets Relation} component that performs the operation $\texttt{FII} \subseteq \texttt{FSI}$.

We were able to verify the frame conditions of all routines in the EiffelBase 2 library. For each routine of each class, the set of class attributes being modified by the routine is indeed a subset of the set of class attributes being named in the postcondition of the routine. In fact, the sets are the same. The results about the process of verifying the library are not surprising -- as the library is fully equipped with contracts, modify clauses and has been fully verified. What it is interesting is that AutoFrame is able to perform the verification automatically.

\subsection{Limitations and Future Work}

Even though the results of our experiments in section \ref{subsection:using} produce the exact same set of modify clauses in EiffelBase2 library, in theory the calculus (and hence its implementation) does introduce potential loss of precision, imposing some limitations on the tool. These are the sources of imprecision
\begin{itemize}
\item the tool treats conditional instructions \eifkw{if} \eif{c} \eifkw{then} \eif{p} \eifkw{else} \eif{q} \eifkw{end} as non-deterministic choice written \eifkw{then} \eif{p} \eifkw{else} \eif{q} \eifkw{end} (it ignores the condition \eif{c}), which executes either \eif{p} or \eif{q}. This is not surprising as, in general, this is an undecidable problem. In practice, the tool could narrow down the cases at least for trivial cases; as a trivial example, ignoring the condition in \eifkw{if} \eif{n $>$ n + 1} \eifkw{then} \eif{a := b} \eifkw{else} \eif{c := d} \eifkw{end} leads to concluding wrongly (that is to say, soundly but with a loss of precision) that both \eif{a} and \eif{c} are in the change set;
\item the tool treats in a similar way the exit condition of loops (and recursion). Another source of imprecision in loops (and recursion) is the use of a fixpoint. As mentioned before, rule \texttt{CC-Loop} uses an upper bound $i$ that in practice is set to $i=3$. This upper bound is a constant but it could be a parameter for the user to tweak. The greater the value of the upper bound the more accuracy, but also the more time to give a result;
\item since AutoFrame treats Dynamic Binding as a non-deterministic choice (as shown before), this is another source of loss precision. The more heirs a class being analyzed have, the more imprecision is introduced.
\end{itemize}
These sources of imprecision open a path to our Future work. We plan to investigate different ways to make the analysis as precise as possible. For instance, we plan to investigate whether different provers could help in the process of determining if a condition can evaluate to true or false statically. 

The tool assumes the command-query separation principle \cite{Meyer:1997:OSC}: asking a question should not change the answer. This cannot be enforced, therefore the calculus might (potentially) yield an incomplete result. As future work, we plan to use AutoFrame to flag those functions that are not pure. Another source of uncertainty (and hence possibly incompleteness) is when the code of the routine being analyzed by the tool is not available. It is common in programming languages to have functions to low-level details (e.g. I/O functions) where code is not available or it is written in another programming language (e.g. C). AutoFrame does not deal with these cases yet. One possible solution is to manually annotate those functions with the corresponding frame conditions. 

Previous subsections evaluate the calculus and its implementation by means of an empirical evaluation. As shown, we have applied the tool to different scenarios. There is not a formal proof of soundness, however. It is important to formally prove that the approach indeed yields (with some imprecision as the problem is undecidable) the right frame conditions. We plan to formally give a proof of soundness of the calculus and its implementation. We plan to take advantage of the fact that the calculus is based on the theory of ``duality semantics'', an application of ideas from Abstract Interpretation \cite{Cousot:1977, Nielson:1999}. We also plan to calculate the complexity of the algorithms used. So far, we based our results in empirical evaluations.


\section{Questioning the benefits}
\label{sec:concl}
In assessing the potential of AutoFrame, we should consider the ``Assertion Inference Paradox'', which \cite{Furia2010} introduced in the following words (abridged):

\begin{adjustwidth}{1cm}{1cm} \textit{``Any  verification  technique  that infers specification  elements from  program  texts faces a risk of vicious circle: the Assertion Inference Paradox.
A program is correct if its implementation satisfies its specification; to  talk  about  correctness  we need both  elements, implementation and specification. But if
we infer the specification from the implementation, does the exercise not become
vacuous?''}
\end{adjustwidth}

The  technique presented in \cite{Furia2010} was for inferring \textit{loop invariants} from program texts, as pioneered by such tools as Daikon \cite{Ernst:01}. Invariant inference has to address the Assertion Inference Paradox: if we infer the specification from the implementation, aren't we just ``documenting the bugs''? \cite{Furia2010} analyzes and answers that objection. For the inference of \textit{frame conditions} as discussed in this article, the risk is much less significant. The frame properties (the specification of what can change) are not typically something that programmers will want to specify explicitly.

In particular, if we want to perform full functional verification, requiring that we work with programs equipped with full contracts (as for EiffelBase 2 with AutoProof, or programs  in JML or Spec\#), the frame conditions involve no surprise: typically, no command, such as \eif{deposit} has an effect on a query unless it lists it explicitly in its postcondition:
\begin{itemize}
\item
If the postcondition talks about a certain query, for example by stating  \eif{balance = } \eifkw{old } \eif{balance} + \eif{sum}, the query can change as a result of executing the command.
\item If it does not list a query, such as \eif{bank}, the query does not change.
\end{itemize}

All full-functional specifications that we have seen satisfy this rule, which \cite{Meyer:Alias:14} analyzes  further under the name ``implicit convention''. It conforms to intuition: if you are going to prove the full functional correctness of a program and state that a query will change, you will also specify \textit{how} it changes. Our experience with proofs of full functional correctness has shown no counter-example: in all formally verified software that we have seen, relying on a formal notation such as JML and Eiffel used with AutoProof that provides \eifkw{modifies} clauses or equivalent, every query appearing in a \eifkw{modifies} clause also appear in the postcondition.

In other words AutoFrame, does not on its own make up any crucial specification property: it simply documents properties that the specifiers would have to write anyway. It removes the tedium and possible errors.

In discussions with us, the authors of AutoProof and of the fully verified EiffelBase 2 library \cite{Polikarpova:14} tended to downplay this benefit, stating that writing the \eifkw{modifies} clauses was ``not such a big deal''. We respect this view, but note that these colleagues are pioneers in software verification. As the verification community strives to make the technology mainstream, it is essential to remove any hurdle that, while not critical for researchers, may turn away a broader audience.

Perhaps even more importantly, we should take into account, as always in software engineering, the role of change. Even if we granted that the initial effort of writing frame clauses is manageable, both the implementations and the specifications will change, making it necessary to update the frame clauses and raising the possibility of errors. Automatic frame inference removes that risk.

\section*{Acknowledgements}
We are indebted to colleagues who collaborated on the previous iterations of the Alias calculus work, particular Sergey Velder for many important suggestions regarding the theory, and Alexander Kogtenkov who implemented an earlier version of the Change calculus. During a talk by one of us (Meyer) at an IFIP WG1.9 meeting in Paris in 2016, Jean-Christophe Filli\^atre uncovered a conceptual oversight of that earlier approach. We thank members of the Software Engineering Laboratory at Innopolis University, particularly Manuel Mazzara and Alexander Naumchev, for many fruitful discussions.

\bibliographystyle{plain}
\bibliography{bibl}
\end{document}